\documentstyle[aps,prl,psfig]{revtex}
\begin{document}

\title{Light Front   Theory Of  Nuclear Matter}
\author{G.A. Miller}
\address{Department of Physics, Box 351560, University of Washington,
Seattle, Washington 98195-1560}
\author{R. Machleidt}
\address{Department of Physics, University of Idaho, Moscow, Idaho 83844}

\date{\today}

\maketitle

\begin{abstract}

A relativistic light front formulation of nuclear
dynamics is applied to infinite nuclear matter.
A hadronic meson-baryon Lagrangian,  consistent with chiral symmetry,
leads to a nuclear eigenvalue problem which is solved,
including nucleon-nucleon NN correlations, in the 
 one-boson-exchange approximation for the NN 
potential. 
The nuclear matter
saturation properties
are reasonably well reproduced, with a compression modulus 
of $180$ MeV. 
We find that there are $\approx 0.05 $ excess pions per nucleon.
\end{abstract}

\pacs{13.75.Cs, 11.80.-m, 21.65.+f, 21.30.-x}

\twocolumn

An important class of high-energy experiments  involving nuclear
targets is most easily understood using 
light-front variables.
Consider  the  lepton-nucleus deep inelastic scattering
 experiments\cite{emc} which found
a significant difference between the parton distributions
of free nucleons and nucleons in a nucleus. This is 
 interpreted as a $\sim$ 10\%
shift in  the momentum distribution of
valence quarks towards smaller values of the Bjorken variable
$x_{Bj}$. The
Bjorken variable is a ratio of the plus-momentum $p^+=p^0+p^3$
of a quark to that 
of the target. Treating the nucleus as a collection of nucleons gives
 $x_{Bj}=p^+/k^+$, where  $k^+$ is the 
plus momentum of a bound nucleon. 
The 
 spatial variable  canonical to $k^+$
 is $x^-=x^0-x^3$, and the time variable is $x^0 +x^3=x^+$.
To use these variables is to use the light front variables
 of Dirac\cite{Di 49}. See the recent reviews\cite{lcrevs}. 

Deep inelastic scattering depends on
the probability 
$f(k^+)$ that a bound nucleon has a momentum $k^+$. Other nuclear reactions,
such as (e,e') and (p,2p) depend also on this very same
probability\cite{fs1fs2}.  The quantity 
$f(k^+)$ is simply related to the square of the ground state wave
 function, provided light front dynamics are used.
 In the
 equal-time
formulation $f(k^+)$ is obtained as a response function that  depends 
on matrix elements between the ground and
all excited states.

 The use of light front
variables is convenient for interpreting   experiments, but
one is faced with  computing the ground state nuclear wave
function using light front 
variables. We
introduce here
a theory of infinite nuclear matter, using light front dynamics,
in which the effects of
correlations are taken into account. This is  a light front
Brueckner theory of nuclear matter.

 Light front dynamics  is a three-dimensional Hamiltonian formalism.
  A  Lagrangian is chosen and one 
derives field equations which
 allow the  elimination of 
dependent degrees of freedom. The light front
Hamiltonian $P^-$  is obtained from
$T^{+-}$ ($T^{\mu\nu}$ is the energy momentum  tensor), and $P^+$, the
plus-momentum
operator is obtained from $T^{++}$. We use  
 a chiral
model, detailed in Refs.\cite{gam97,mm99}, in which the 
nuclear constituents are nucleons $\psi$, pions, 
 scalar mesons,
 vector mesons,   and $\eta$ mesons.

We start by briefly reviewing some of the necessary formal details
presented in Ref.~\cite{gam97} (which presented a mean field 
calculation of the properties of nuclear matter). Our specific
equations employ the rest frame of nuclear matter. 
The derivation of $T^{+-}$ is accomplished by solving for  the
dependent nucleon fields in terms of the independent ones. The
simple completion of this task requires setting the $^+$
components of the vector fields $V^\mu$
 to zero.  This is accomplished using a
transformation \cite{soperyan} on the nucleon fields, in which 
one replaces $V^\mu$ with other fields $\bar{V}^\mu$
($\bar{V}^+=0$) which enter in the nucleon  field equations.
 We carry this feature  through in our calculations (including the
 related mean fields), but
drop the bar to simplify the notation.
More detailed explanations are given in Refs.\cite{gam97} and
\cite{mm99}.

We  now confront  the eigenvalue problem for 
the ground state of infinite nuclear matter: 
\begin {equation}
P^-|\Psi\rangle=M_A|\Psi\rangle. \label{sc}
\end{equation} 
 For a nuclear system at rest we must also satisfy 
$
P^+|\Psi\rangle=M_A|\Psi\rangle. \label{scp}
$
The operator $P^-$ can be divided into kinematic
 $P^-_0(N)$ and interaction $J$ terms:
\begin{equation}
 P^- = P^-_0(N) + J \label{op}
\end{equation}
in which $P^-_0(N)$ is the kinetic contribution to the $P^-$ operator,
 giving ${p_\perp^2 + m^2}\over{p^+}$ for the minus-momentum of free
nucleons.
The operator $J$ is the sum of three terms
$
J \equiv v_1 + v_2 + v_3:$ 
$v_1 $ gives the single meson-nucleon vertex functions;
$v_2 $ accounts for  meson emission followed
by instantaneous nucleon  propagation 
followed by another meson emission; and, $v_3 $ accounts for the 
instantaneous propagation of vector mesons.

The
traditional path of using a two-nucleon potential and temporarily
eliminating  the meson degrees of freedom by
adding  and subtracting  a  two-nucleon potential
to the Hamiltonian is followed. The 
terms involving the difference between $J$ and
the two-nucleon potential are handled later using   perturbation theory.
Then the  effective Lagrangian is 
$
 {\cal L_K} \equiv\bar{\psi}(i\gamma\cdot\partial -M)\psi -{{{\cal K}\over 2}},
$
with 
the two-nucleon interaction density ${{\cal K}}/2$.
Using ${\cal L_K}$, gives 
the corresponding $P^\pm$ operators, with  
$
P^-_K 
=P^-_0(N) +K$.
 $K$ is the volume integral of ${{\cal K}}$,
and 
\begin{eqnarray}P^- &=& P^-_K +J-K +P^-_0(m), 
\label{full} \end{eqnarray} 
where $P^-_0(m)$ accounts for the non-interacting mesonic contribution
to $P^{-}$.
 
The purely nucleonic   part of the full wave function is defined as 
$\mid \Phi  \rangle$, and is
the solution of the light front Schroedinger equation
\begin{equation} P^-_K \mid \Phi \rangle
  =\left( P^-_0(N) +K\right)\mid \Phi \rangle=
M_0 \mid \Phi \rangle  .\label{big}
\end{equation}

The interaction $K$ is obtained
using 
second-order perturbation theory. Consider 
the scattering process $1+2\to 3+4$. We find 
     \begin{equation}\langle 3,4|
 K|1,2\rangle=\langle 3,4| v_1g(P_{ij}^-)v_1+v_3|1,2\rangle,
 \label{kdef0}
 \end{equation}
 with
$
  g_0(P_{ij}^-)\equiv {1\over P_{ij}^--P_0^-}\;,  
  $
  where $P_{ij}^-$ is the negative component of the total initial (12)
or final (34)
  momentum which here  are the same.

We discuss  the energy denominator $(P_{ij}^--P_0^-)k^+$.
 Suppose that $k_1^+>k_3^+$. Then
 the emitted meson of mass $\mu$
 and momentum $k$, with  
$k^+=k_1^+-k_3^+, \bbox{k}_\perp=
 {\bbox{k}_1}_\perp-{\bbox{k}_3}_\perp$, $k^-={k_\perp^2+\mu^2\over k^+}$,
 and
 $  k^+( P_{ij}^--P_0^- )= k^+(k^-_1-k_3^-)
    -{{k_\perp}^2+\mu^2} =q^2-\mu^2,
    $ 
    where $q\equiv k_1-k_3$. That this 
    form is the Lorentz-invariant inverse of the Klein-Gordon propagator
 indicates that $K$ accounts for the nucleon-nucleon
 one-boson exchange potential. 
Note that  $q^2={q_0}^2-\vec{q}^2$, with  ${q_0}^2$ accounting
for  retardation
effects.

The interaction $K$ is strong,
so that  we sum the unitary set of iterations  to
obtain 
the two-nucleon integral equation for the transition matrix $T$:
\begin{eqnarray} 
\langle3,4|
T|1,2\rangle
&=&\langle 3,4|
V|1,2\rangle+
\int\sum_{\lambda_5,\lambda_6} \langle 3,4|
V|5,6\rangle 
\nonumber \\
 & & \times
{2M^2\over \alpha(1-\alpha)}
{d^2p_\perp d\alpha\over P_{ij}^2-{p_{\perp}^2+M^2\over\alpha(1-\alpha)}
+i\epsilon}
\langle5,6|T
|1,2\rangle. \label{419}
\end{eqnarray}
The plus-momentum   variable is expressed in terms of a 
momentum  fraction
$\alpha$ such that
$
p_5^+=\alpha P_{ij}^+,
$
and $
\bbox{p}_\perp\equiv (1-\alpha)\bbox{p_5}_\perp-\alpha
\bbox{p_6}_\perp  
.$ 
The potential $V$ is a kinematic factor times $K$. 

We mention again that  $V$ is derived using the one
boson exchange approximation, so that effects of 
multi-mesons in flight are not included in Eq. (6). 
The effects of the stretched box diagram
has been studied  in the light front calculations
of Ref.~\cite{nico}, and can be 
important in
maintaining the full Lorentz invariance of the scattering amplitude.
The present results (6) for the on-shell $T$-matrix
are invariant under boosts in the
$z$-direction.

The Weinberg-type\cite{fs1fs2} equation (\ref{419}),  can
be re-expressed 
using the variable transformation\cite{Te 76}:
$
\alpha={E(p)+p^3\over 2E(p)} \label{alpha}
$, 
with $E(p)\equiv\sqrt{\bbox{p}\cdot\bbox{p}+M^2}$.
The result is:
\begin{eqnarray}
\langle3,4|T
|1,2\rangle
&=&\langle3,4|V
|1,2\rangle+\int
\sum_{\lambda_5,\lambda_6} \langle 3,4|V
|5,6\rangle
\nonumber \\
 & & \times
{M^2\over E(p)}
{d^3p \over p_i^2-p^2
+i\epsilon}
\langle5,6|T
|1,2\rangle, \label{bsbs}
\end{eqnarray}
which shows a manifest rotational invariance. 


The one-boson exchange
potential $V$ obtained with our formalism is constructed
elsewhere\cite{mm99}.
We choose the meson-nucleon coupling constants and include 
the effects of form factors to  reproduce the NN data.
This light front work 
yields potentials and transition matrix elements 
essentially identical to the Bonn OBEP potentials
\cite{Mac89,Mac93} because the  differences arising  from including 
retardation effects, and using Eq.~(\ref{bsbs}) 
are very small. Thus we find
a good description of the deuteron
and the two-nucleon scattering data below the
inelastic threshold\cite{mm99}.

Now that $V$ is determined, we turn to solving
Eq.~(\ref{big}). It is worthwhile to begin by explaining
 how our light front quantized
calculation of the energy and density
of nuclear matter  includes  the effects of
two-nucleon NN correlations. We   employ a light front version of the
standard technique of 
obtaining
 the  G-matrix  (see Eqs. (11) and (12) below)
which is a 
nuclear matter generalization of 
the  T-matrix of Eq. (6). 
We then determine the  nuclear density by finding  the
Slater determinant $|\phi>$ which minimizes the expectation value of
$P^-$ subject to the constraint that the expectation values of $P^+ $
and $P^-$ are equal. The single particle wave functions that make up
the determinant $|\phi>$ are to be given in Eq. (9), which is not the Dirac
equation.  The self-consistent fields $U$ which enter  are
to be obtained from the expectation value of the G-matrix in
$\mid\phi\rangle$. The use of symmetries\cite{bdsjdw}
causes the 
only non-vanishing mean-fields to be scalar  $U_S$
and vector fields $U_V^\mu$. Neglect of the very small \cite{bdsjdw}
spatial component of $U_V^\mu$ leaves us with  $U_V^-$. 

We now obtain the Slater determinant $\mid \phi\rangle$ and the 
related 
scalar $U_S$ and vector $U_V^-$
potentials. The eigenvalues
and eigenvectors of the resulting light-front mean field Hamiltonian
$P^-_{MF}$ are easy to obtain,  and can be chosen to best approximate the
effects of the two-nucleon interaction. The mean field
light front Hamiltonian density $T_{MF}^{+-}$ is obtained 
and its volume integral 
is a single-nucleon operator  $P^-_{MF}$. If one expands $\psi$ 
in terms of the single particle wave equation including
$ {U}$  as the only interaction, one finds that
$P^-_{MF}= P_0^-(N).
$ 
  The  ground state
eigenvector of this operator  is a Slater
determinant denoted as $\mid\phi\rangle$:  
$  P_0^-(N)\mid\phi\rangle
= m_0\mid\phi\rangle, $ and 
\begin{eqnarray}
  \mid\Phi\rangle=\mid\phi\rangle+{1\over M_0-P_0^-(N)-K}\Lambda K
\mid\phi\rangle,\label{phi}
\end{eqnarray} 
where $\Lambda \equiv 1 - \mid \phi \rangle \langle \phi \mid .$
The single nucleon orbitals $\mid i\rangle$ are obtained from
\cite{gam97}:  
\begin{eqnarray}
\left(P^-_{MF}-U_V^-\right)\mid i\rangle_+
&=& \left(\bbox{\alpha}_\perp\cdot\bbox{k_i}_\perp +\beta (M+U_S)\right)
\mid i\rangle_+ ,\nonumber\\
k^+\mid i\rangle_-&=& \left(
\bbox{\alpha}_\perp\cdot\bbox{k_i}_\perp+\beta(M+U_S)\right)\mid
i \rangle_+,\nonumber\\
\left(P^-_{MF}-U_V^-\right)\mid i\rangle_+
&=& {{\bbox{k}_i}_\perp^2 +(M+U_S)^2\over k^+_i}\mid i\rangle_+
.\label{pim}
\end{eqnarray}
The subscripts $\pm$ refer to the independent ($+$) and dependent ($-$)
projections of the nucleon fields.

Using 
standard manipulations on Eq.~(\ref{phi}), and keeping correlations between
pairs of nucleons only 
leads to: 
\begin{eqnarray}
M_0 - m_0 \approx \langle \phi \mid
{1\over2}\sum_{i,j} \Gamma_{i,j}(P^-_{ij})\mid \phi\rangle, 
 \label{mass}
\end{eqnarray}
where $\Gamma_{i,j}$ is a two-nucleon operator:
\begin{eqnarray}
\Gamma_{i,j}(P^-_{ij}) &=& K_{ij} + K_{ij} \,\,\Lambda{1\over
 P^-_{ij}-\Lambda P_0^-(N)\Lambda}\Lambda \,\, \Gamma_{i,j}(P^-_{ij}).
\label{gamma}\end{eqnarray}
The notation $i,j$ refers to the nucleon orbitals of Eq.~(\ref{pim}). 
The approximation that $U$ 
 is independent of orbital $i$ causes the vector 
potential to cancel out in  the denominators e.g. $ P_{ij}^--(p_5^-+p_6^-)$.
Starting with Eq.~(\ref{gamma}), we obtain  
an equivalent three-dimensional, manifestly rotationally invariant 
 integral equation 
by using 
 the  medium-modified version of the
variable transformation\cite{Te 76}:
$
\alpha={E^*_{\bbox{p}}+p^3\over 2E^*_{\bbox{p}}},
$
with $E^*_{\bbox{p}} \equiv\sqrt{\bbox{p}\cdot\bbox{p}+{M^*}^2}$.
 The result is:
\begin{eqnarray}
\langle3,4|G
|1,2\rangle
&=&\langle3,4|V
|1,2\rangle+\int
\sum_{\lambda_5,\lambda_6} \langle 3,4|V
|5,6\rangle
\nonumber \\
 & & \times
{{M^*}^2\over E^*_{\bbox{p}}}
{d^3p\; Q \over p_i^2-p^2}
\langle5,6|G
|1,2\rangle, \label{bbsbs}
\end{eqnarray}
where $G$ is related to $\Gamma$ by a kinematic factor. 
The operator $Q$  is the two-body version of
$\Lambda$ and
projects the momenta $p_5$ and $p_6$ above the
Fermi sea.
The evaluation of $V$ using the basis
of  Eq.~(\ref{pim})
is the essential difference between relativistic and non-relativistic
Brueckner theory.

We now specify
 the Slater determinant $\mid\phi\rangle$.
The occupied states are to 
 fill up a Fermi sea, defined in terms of a Fermi momentum,
  $k_F$, that 
is the magnitude of a three vector. This three vector can be 
defined\cite{gs} as:
$
k^+=E^*({\bbox k})+k^3$, 
which specifies $k^3$. 
Computing the energy and plus 
momentum  then  proceeds by  taking the  expectation
values of $T^{+-}$ and $T^{++}$. 
Note  that 
the resulting  formulae can be cast in a more familiar form
by using a discrete
representation of the single nucleon states, 
  a set of spinors:
$\mid\alpha\rangle$, with $\alpha \equiv \bbox{k},\lambda$ such that
$\langle\alpha\mid\alpha\rangle=1$. These light front spinors of
Eqs.~(9) can be related to equal time spinors. Multiply the first of (9)
by $\gamma^+$ and the second by $\gamma^-$. Then add the two equations
to obtain a new equation. Its
solution turns out to be  a spinor 
$\langle x\mid\alpha\rangle$  which differs from the equal time
(ET)
spinor by only a phase
factor,
$\exp{( -i{U_V^0x^-\over 2})}$,
which cancels out of matrix elements under our approximation that
$U_V^-$ is  a constant. Then 
the Brueckner light front Hartree-Fock (BHF)
 approximation is to the mean-field is given by 
\begin{eqnarray}
   U(\alpha)
&=&
 \sum_{\beta<F} Re
   \langle\bar{\alpha}\bar{\beta}\mid G
   \mid\alpha\beta\rangle_a. \label{uhat} 
\end{eqnarray}

The nuclear matter $G$-matrix is the solution of the integral
equation,
(\ref{bbsbs}) in a plane wave basis.
 We use the implicit definition of $k^3$ to obtain 
the total $\bbox{P}$ and relative three-momenta of two nucleons
interacting in nuclear matter. An angle averaging procedure ~\cite{HT70},
is used for ${\bbox P}$ and to obtain the 
Pauli operator $Q$.
The calculation of the nuclear matter $G$ matrix involves a self-consistency,
since the solution of Eq.~(\ref{bbsbs}) for $G$ requires knowledge
of $M^*$ which, in turn, is determined from $G$ via 
Eq.~(\ref{uhat}). 
Solutions are obtained using an iterative procedure.
We obtain the nuclear mass from  
 the average\cite{gam97} of  $P_A^\pm$:
$
M_0=\sum_{\alpha<F} \epsilon_\alpha
   +{1\over 2}\sum_{\alpha,\beta<F}
   \langle\bar{\alpha}\bar{\beta}\mid G
   \mid\alpha\beta\rangle_a. \label{main}
$ 

The results for 
the binding energy per nucleon $E/A$
in nuclear matter as a function of density,
are shown in Fig.~1 as the solid line.
The curve has a minimum  at ${E}/A = -14.71$ MeV and $k_F = 1.37$
fm$^{-1}$, and the  incompressibility (compression modulus) is 
$K=180$ MeV at the minimum.
These values  agree well with the
empirical values 
${E}/A=-16\pm 1$ MeV,
$k_F=1.35\pm 0.05$ fm$^{-1}$, 
and $K=210\pm 30$ MeV~\cite{Bla80}.
Note also that the effect of using non-relativistic Brueckner theory is
that the saturation density is predicted too high, but the incompressibility
is  about the same as ours\cite{bhf}.
Comparison with the results of  an earlier relativistic
calculation~\cite{rm1} shows 
that 
the only significant difference occurs for the incompressibility
which is reduced from the earlier value of $250$ MeV.
This is due to using medium-modified  single-particle energies  
in the evaluation of the retardation term\cite{excuse}, and also 
to using  new values of the mean fields:
$M^*(k_F)=M_N+U_S=718 
$ MeV and $U_V=164.4 $ MeV.
These potentials have magnitudes that are 
considerably smaller than those obtained using
mean field theory\cite{foot3}.

\begin{figure}[h]
\vspace*{1cm}
\psfig{figure=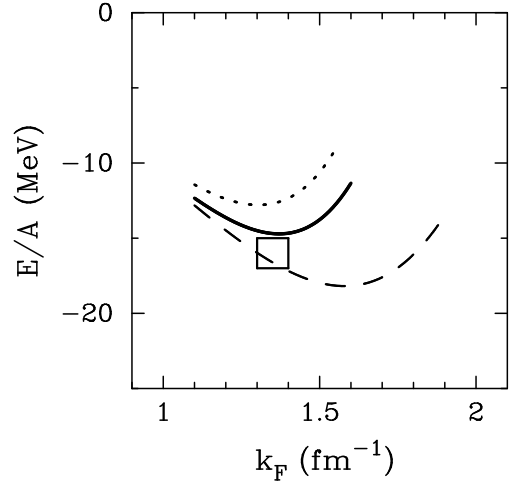,height=8cm}
\vspace*{-2cm}
\caption{Binding Energy per nucleon in nuclear matter, E/A 
as a function of Fermi momentum $k_F$.
The solid line is our result using
light-front Brueckner theory. The dotted curve is obtained
when the medium effect on meson retardation is omitted.
The dashed line is the result from nonrelativistic
Brueckner theory. The box represents the range of values allowed by 
extrapolation from data.}
\end{figure}

Turn now to the mesonic component of the nuclear  wave function.
The purely nucleonic part of the Fock space is $\mid \Phi \rangle $ of
Eq.~(\ref{phi}), and 
the full wave function is 
 $\mid \Psi \rangle $ of Eq.~(\ref{sc}). 
The mesonic content is contained in the difference between the two
wave functions.
Using Eq.~(\ref{full}) in Eqs.~(\ref{sc}) and (\ref{big}) allows us 
to  obtain 
   \begin{equation}
 \mid \Psi \rangle = \mid \Phi \rangle
 + {1 \over M_A- \Lambda_\Phi P^- \Lambda_\Phi}
 \Lambda_\Phi 
(J-K) \mid \Phi\rangle , \label{psi}
\end{equation}
where $\Lambda_\Phi=1-\mid\Phi\rangle\langle\Phi\mid$.
The second term accounts for the mesonic content.
Taking the expectation value of the mesonic number operators
yields the average number of mesons $N_m$: 
\begin{eqnarray}
N_{m}
\approx 
 -\langle\Phi \mid
 {\partial V_{m}(P^-_{ij}) \over \partial P^-_{ij} }
 \mid \Phi \rangle 
\end{eqnarray}
where $V_m$ represents the potential caused by the exchange of
a pion or scalar meson, and its dependence on   $P^-_{ij}$ is given in
Eq.~(\ref{kdef0}).
For the pion the result is  about 5.0\% per nucleon 
\cite{good}. This is smaller than the 18\% of Friman et al\cite{bf}
because we use scalar mesons  instead of intermediate $\Delta$ states to 
 provide the bulk of the attractive 
force.

The values of $U_S$, $U_V^-$ and $N_\pi$ allow us to assess the 
deep inelastic scattering of leptons from our version of the ground state of
nuclear matter. Using $M^*(k_F)=744$ MeV~\cite{foot3} and 
neglecting the influence of
two-particle-two-hole states
to approximate
$f(k^+)$\cite{gam97} 
shows that nucleons carry 81\% (as opposed to the 65\% 
of mean field theory\cite{gam97}) of  the nuclear plus momentum. 
This represents a vast improvement in the description of nuclear deep inelastic
scattering. The 
minimum value of the ratio $F_{2A}/F_{2N}$, obtained from 
the convolution formula
\begin{equation}
{F_{2A}(x)\over A}=\int_x^A dy f(y) F_{2N}(x/y), \label{deep}
\end{equation}
where $y=A k^+/(M_N+E/A)$, 
and
$x=x_{Bj}M_N/M_A$, is increased by a factor of twenty   towards 
the data 
as  
extrapolated in Ref.~\cite{Sick}. But this calculation provides only  a
lower limit of the nucleon contribution because of the neglect of 
effects of the two-particle-two hole states~\cite{foot4}.

Turn now to the experimental information about the nuclear pionic content.
The Drell-Yan experiment on nuclear targets\cite{dyexp}
showed no enhancement of nuclear pions within an error of about 5\%-10\% for
their heaviest target. No substantial  pionic enhancement 
is found in (p,n) reactions\cite{pn}.
Understanding this result is 
an important challenge to the 
understanding of nuclear dynamics~\cite{missing}. 
Here we have a good description of nuclear dynamics, 
and our 5\%  enhancement is consistent\cite{jm},
within errors, with the Drell-Yan
data.

This paper contains a new 
relativistic light-front theory of nuclear matter which
leads to a good
description of the binding energy, density, and 
incompressibility of nuclear matter.
The use of a meson-nucleon Lagrangian enables us to also compute the 
mesonic content of the  wave function using a consistent approach represented
by Eqs.~(\ref{sc}),(\ref{full}),(\ref{big}) and (\ref{psi}).
  The results are not in conflict
with extrapolations of deep inelastic scattering and Drell-Yan data to 
nuclear matter. 

This work was supported in part the the U.S.\ National Science
Foundation under Grant No.\ PHY-9603097., and by the U.S.D.O.E.
We acknowledge useful conversations with B.L.G. Bakker.


\begin{references}
\bibitem{emc}  J. Aubert {\it et al.,} Phys. Lett. {\bf B123}, 275 (1982).
 R.G. Arnold,  {\it et al.,}Phys. Rev. Lett. {\bf 52} 727 (1984).  
A. Bodek,  {\it et al.,}Phys. Rev. Lett. {\bf 51, } 534 (1983). 
R.L. Jaffe, 
in Relativistic Dynamics and Quark-Nuclear Physics. 
    Ed. by M. B. Johnson and A. Picklesimer;  
M. Arneodo, Phys. Rep.
240, 301 (1994);
D.F. Geesaman, 
K. Saito, A.W. Thomas, Ann. Rev. Nucl. Part. Sci. {\bf 45}, 337 (1995).
  

\bibitem{Di 49}  
P.\, A.\, M.\, Dirac, 
Rev. Mod. Phys. {\bf 21}, 392 (1949).
\bibitem{lcrevs}
S. J. Brodsky, H-C Pauli,
S.\,S.\, Pinsky, 
Phys.Rept. {\bf 301}, 299,(1998);
M.  Burkardt Adv. Nucl. Phys. {\bf23}, 1 (1993);
A. Harindranath,
 hep-ph/9612244. 
\bibitem {fs1fs2} L.~L. Frankfurt and M.~I. Strikman 
Phys. Rep. {\bf 76}, 215 (1981);
L.L. Frankfurt and
 M.I. Strikman, Phys. Rep.  {\bf 160}, 235 (1988).
\bibitem{gam97}G.A. Miller, Phys. Rev. {\bf C56}, R8 (1997); 
G.A. Miller, Phys. Rev. {\bf C56}, 2789 (1997).
\bibitem{mm99} 
G.A. Miller and R. Machleidt,
UW preprint {NT@UW-98-35}, 
in preparation.
\bibitem{soperyan} D.E. Soper, 
SLAC pub-137 (1971); 
D.E. Soper, Phys. Rev D4, 1620 (1971);
T-M Yan  Phys. Rev. {\bf D7}, 1760, 1780 (1974).
\bibitem{nico}
 N.C.J. Schoonderwoerd, B.L.G. Bakker, V.A. Karmanov,
Phys. Rev.{\bf C58}, 3093 (1998); 
N.C.J. Schoonderwoerd, B.L.G. Bakker, Phys.Rev.{\bf D57}, 4965 (1998);
N.C.J. Schoonderwoerd, B.L.G. Bakker,
Phys.Rev.{\bf D58}, 025013 (1998).

\bibitem{Te 76} M.V. Terent'ev, Sov. J. Nucl. Phys. {\bf 24}, 106 (1976).
\bibitem {BbS} R. Blankenbecler and
  R. Sugar, Phys.Rev. {\bf 142}, 1051 (1966).
\bibitem{Mac89}
R. Machleidt,  Adv. Nucl. Phys. {\bf 19}, 189 (1989).
\bibitem{Mac93} R. Machleidt, {\it One-Boson-Exchange Potentials
and Nucleon-Nucleon Scattering}, in: {\it Computational Nuclear Physics 
2---Nuclear Reactions}, K. Langanke, J. A. Maruhn, and S. E. Koonin,
eds.\ (Springer, New York, 1993), Chapter~1, p.~1.

\bibitem{bdsjdw} B.D. Serot and J.D. Walecka, Adv. Nucl. Phys. {\bf 16},1 (1986)
\bibitem{gs} St. Glazek, C. M. Shakin, Phys. Rev. C44, 1012 (1991).
\bibitem{HT70} M. I. Haftel and F. Tabakin, Nucl. Phys. 
{\bf A158}, 1 (1970).


\bibitem{Bla80} J.P. Blaizot, Phys. Rep. {\bf 65}, 171 (1980).

\bibitem{bhf} P. Grang\'{e} and M.A. Preston, Nucl. Phys.{\bf A204}, 1 (1973);
  Nucl. Phys.{\bf A219}, 266 (1974);
  M. Saraceno, J.P. Vary, G. Bozzolo and H.G. Miller,
  Phys. Rev. {\bf C37}, 1267
(1988).

\bibitem{rm1} R. Brockmann and  R. Machleidt,
Phys. Lett. {\bf 149B}, 283 (1984); Phys. Rev. C {\bf 42}, 1965 (1990);
in  ``Open Problems in Nuclear Matter'', M. Baldo, ed.  
(World Scientific, Singapore, 1998) 
nucl-th/9612004. 

\bibitem{excuse} 
The effects of the medium  on the retardation inherent in 
 the meson-nucleon form factors are ignored
 because the mean field should vanish at  high  momentum.
 
\bibitem{foot3} $U$ depends on momentum, so to use 
 a constant value is to take an average.
 Eqs.~(\ref{bbsbs}) and (\ref{uhat})
are non-linear, and the solutions have 
a  spread of values. Indeed, 
 $M^*(k_F)= 744$ also  leads to good saturation properties.

\bibitem{good} The relatively small value of $N_\pi$ indicates that 
the effects of $J-K$ in Eq.~(\ref{full}) are small, and that $M_0$ 
 is a good approximation to $M_A$. 
\bibitem{foot4} We estimate  that
nucleons with momentum greater than $k_F$ would substantially increase the
computed ratio $F_{2A}/F_{2N}$ because $F_{2N}(x)$ decreases very rapidly
with increasing values of $x$ and because $M^*$ would increase at high momenta.
\bibitem{bf} B. Friman, V.R. Pandharipande, and R.B. Wiringa Phys. Rev. Lett
  {\bf 51}, 763 (1983)


\bibitem{Sick} I. Sick and D. Day, Phys.Lett. {\bf B274}, 16 1992).
     
\bibitem{dyexp}D.M.  Alde et al. Phys. Rev. Lett. {\bf 64}, 2479 (1990).
\bibitem{pn} 
  L. B. Rees et al., Phys. Rev. C 34, 627 (1986); 
    J. B. McClelland et al., Phys. Rev. Lett. 69, 582 (1992); 
    X. Y. Chen et al., Phys. Rev. C 47, 2159 (1993); 
   T. N. Taddeucci et al., Phys. Rev. Lett. 73, 3516 (1994). 
\bibitem{missing} G.F. Bertsch, L. Frankfurt, and M. Strikman,
 Science  {\bf 259}, 773 (1993).
 \bibitem{jm} H. Jung and G.A. Miller, Phys. Rev. {\bf C41},659 (1990).
\end{references}
\end{document}